\documentclass[pra,aps,superscriptaddress,showpacs,floatfix,tightenlines,twocolumn]{revtex4-1}
\usepackage{graphicx}
\usepackage{dcolumn}
\usepackage{bm}
\usepackage{amssymb}
\usepackage{amsmath}
\usepackage{url}
\usepackage{amsthm}
\usepackage{layout}
\usepackage{epsfig}
\usepackage{graphicx}
\usepackage{epstopdf}
\usepackage{booktabs}
\usepackage{float}
\usepackage{subfigure}
\usepackage[hyperindex,breaklinks]{hyperref}

\begin{document}

\title{Detected the steerability bounds of the generalized Werner states via BackPropagation neural network}
\author{Jun Zhang}
\email{zhang6347@163.com}
\affiliation{College of Data Science, Taiyuan University of Technology, Taiyuan 030024, China}
\author{Kan He}
\email{hekanquantum@163.com}
\affiliation{College of Mathematics, Taiyuan University of Technology, Taiyuan, 030024, China}
\affiliation {College of Information and Computer Science, Taiyuan University of Technology, Taiyuan, 030024, China}
\affiliation {College of software, Taiyuan University of Technology, Taiyuan, 030024, China}
\author{Ying Zhang}
\affiliation {College of Information and Computer Science, Taiyuan University of Technology, Taiyuan, 030024, China}
\author{Yu-yang Hao}
\affiliation {College of software, Taiyuan University of Technology, Taiyuan, 030024, China}
\author{Jin-chuan Hou}
\affiliation{College of Mathematics, Taiyuan University of Technology, Taiyuan, 030024, China}
\author{Fang-Peng Lan}
 \affiliation {College of Information
and Computer Science, Taiyuan University of Technology, Taiyuan,
030024, China}
\author{Bao-Ning Niu}
\email{niubaoning@tyut.edu.cn} \affiliation {College of Information
and Computer Science, Taiyuan University of Technology, Taiyuan,
030024, China}
\date{\today }
\begin{abstract}
We use error BackPropagation (BP) neural network to determine whether an arbitrary two-qubit quantum state is steerable and optimize the steerability bounds of the generalized Werner state. The results show that no matter how we choose the features for the quantum states, we can use the BP neural network to construct several models to realize high-performance quantum steering classifiers compared with the support vector machine (SVM). In addition, we predict the steerability bounds of the generalized Werner states by using the classifiers which are newly constructed by the BP neural network, that is, the predicted steerability bounds are closer to the theoretical bounds. In particular, high-performance classifiers with partial information of the quantum states which we only need to measure in three fixed measurement directions are obtained.
\end{abstract}

\maketitle


\section{Introduction}

In 1935, Einstein, Podolsky and Rosen (EPR) questioned the completeness of quantum mechanics based on EPR paradox in their EPR paper \cite{EPR}. The EPR argument led to long lasting discussions. Schr\"{o}dinger introduced the concept of quantum steering \cite{Schrodinger} in order to formalize the ``spooky action at distance" in the EPR paper. In a quantum steering scenario, Alice can steer Bob's quantum state by choosing a proper local measurement. This work did not cause widespread concern. It was not until 2007 that Wiseman, Jones and Dougherty proposed a precise definition and systematic criteria \cite{Wiseman} that people began to pay attention to steering. In the modern view, steering \cite{steer} is a concept incompatible with the local hidden state (LHS) model. It is a quantum correlation between entanglement \cite{entangle1,entangle2} and Bell nonlocality \cite{nonlocal}. Now, steering has been applied to various quantum information processing tasks \cite{Wiseman,app2,app3}, such as one-sided device-independent quantum key distribution \cite{Branciard,qkd2,qkd3,qkd4,qkd5,qkd6}, channel discrimination \cite{Piani,channel}, randomness certification \cite{Passaro,random1,random2} and teleamplification \cite{He}.

The set of EPR steerable states is a proper subset of entangled states, namely, the quantum entangled states are not necessarily steerable quantum state, but the steerable quantum state must be a quantum entangled state. Thus, it is important to detect the steerability of a given state shared by Alice and Bob \cite{Wiseman,app2,app3,Piani,13,14,15,16,17,18,19,20,22,21,30,23,24}. So far, various criteria and inequalities for steering have been proposed \cite{13,14,15,16,17,18,19,20,22,24}. But sometimes it is difficult to determine whether an arbitrary unknown two-qubit quantum state is steerable through these criteria and inequalities. Fortunately, these criteria can be computed by a semidefinite programming (SDP) \cite{SDP} where we must find Alice's optimal measurement direction for a given state shared by Alice and Bob in order to determine whether Alice can steer Bob's state. But this requires a lot of measurements, it becomes very hard to calculate as the number of Alice's measurements increases \cite{steer}. In addition, the situation gets worse when dealing with the steerability of a series of different rapidly generated states.

Machine learning can quickly make predictions on given new data with reasonable accuracy by learning from a large amount of existing data. Recently, machine learning has been applied in quantum physics, such as entanglement \cite{engtanglement}, nonlocality \cite{nonlocality1,nonlocality2}, phase transitions identification \cite{phase1,phase2}, quantum state tomography \cite{Tomolography}, Markovianity \cite{Markovianity} and steering \cite{32,30}. Compared with the SDP method, machine learning requires much less resources and can quickly determine whether a quantum state is steerable. Ren and Chen proposed a machine learning method to detect the steerability of an arbitrary two-qubit state based on support vector machine (SVM), gave the validity and efficiency of steering classification \cite{30}. As we know, SVM maps the samples from the original space to a higher-dimensional feature space through the kernel function, so that the samples are linearly separable in this feature space and the global optimal solution can be found. However, when predict the steerability bounds of the generalized Werner states, the quantum steering classifiers trained by SVM will misjudge the steerable state to be the unsteerable state, vice versa. The main reason is that the data generated by SDP existing errors, that is, based on the finite values of measurements, SDP program for a given quantum state does not necessarily mean that the quantum state is steerable, if we try more measurements and more values of the measurements, the SDP program may conclude that the quantum state is unsteerable.

Thus, we explore a machine learning method using BackPropagation (BP) neural network to reduce the error rate of misjudge, though the misdata generated by SDP can not be avoid. The BP neural network has the advantage of easily adjusting the parameters. Moreover, the BP neural network has a better performance when processing large amounts of data and continuous data features. So no matter what the features of the quantum states are, the BP neural network can realize high-performance quantum steering classifiers compared with the SVM method, namely, it can realized the higher accuracy of classification. In addition, we can use the BP neural network to construct several new classifiers to optimize the steerability bounds of the generalized Werner state. Although the steerability bounds of the generalized Werner state obtained by SVM are better than that obtained by SDP, they are not very close to the theoretical bounds. The steerability bounds of the generalized Werner states in terms of the BP neural network are closer to the theoretical bounds than the SVM approach.

This article is organized as follows:
In Sec. \ref{sec:2}, we introduce some basic concepts, principles and algorithms to be used in this article.
In Sec. \ref{sec}, we illustrate the collection process of our datasets in detail, give several classifiers which are constructed by the BP neural network based on four features, and compare the performance of these four classes of classifiers with the classifiers trained by the SVM method. We also obtain a better steerability bounds of the generalized Werner state. Finally, we summarize our results in Sec. \ref{sec4}.

\section{Preliminaries}\label{sec:2}
\subsection{Quantum steering}
Firstly, we introduce the concept of quantum steering briefly. Assuming that for an unknown quantum state $\rho$ shared by Alice and Bob, Alice performs the quantum measurements $\{M_{x}=\{M_{a|x}\}\}$ on her subsystems and the corresponding measurement outcomes are denoted by $a$. Based on the quantum theory, according to choosing quantum measurement $M_{a|x}$ and the measurement outcome $a$, the unnormalized quantum states assemblage of the subsystem $\rho_{B}$ is $\{p(a|x),\rho_{a|x}\}$
where
\begin{equation}
\rho_{a|x}=\mathrm{tr}_{A}[(M_{a|x}\otimes I^{B})\rho],
\end{equation}
and the probability distribution
\begin{equation}
p(a|x)=\mathrm{tr}[(M_{a|x}\otimes I^{B})\rho].
\end{equation}

Quantum steering describes a scenario: Bob can full control of his measurements and access the condition state $\rho_{a|x}$ without characterization of Alice's measurements. Namely, Bob performs tomography to reconstruct the set of conditional assemblage $\{\rho_{a|x}\}$ and all the results do not dependent on any particular information of how Alice's measurements work. In other words, Alice can steer Bob's local state by performing local measurement and classical communication on the particle she owns. Utilizing the LHS model, quantum steering can be defined as the impossibility of remotely generating ensembles which produced by a LHS model. That is, suppose that a source sends a classical message $\lambda$ to Alice and a corresponding state $\sigma_{\lambda}$ to Bob. If the measurement applied by Alice is $x$, the classical variable $\lambda$ leads to the probability of she getting measurement outcome $a$ is $p(a|x,\lambda)$. Whilst the probability distribution of the classical message $\lambda$ is $p(\lambda)$, since Bob can not access to the classical variable $\lambda$, the finial corresponding assemblage Bob observes is composed by the elements

\begin{align}\label{2}
\rho_{a|x}=\int \mathrm{d}\lambda p(\lambda)p(a|x,\lambda)\sigma_{\lambda}.
\end{align}
If there exists an conditional assemblage $\rho_{a|x}$ corresponding to the quantum state $\rho$ can be generated from a LHS model, then Alice can not steer Bob's state. Otherwise, quantum state $\rho$ is steerable from Alice to Bob. As an example, the generalized Werner state is identified as a simple family of one-way steerable two-qubit states \cite{31}. It can be given by
\begin{align}\label{5}
\rho(\alpha,\chi)=\alpha|\psi_{\chi}\rangle\langle\psi_\chi|+(1-\alpha)\frac{I^{A}}{2}\otimes\rho^{B},
\end{align}
where $|\psi_{\chi}\rangle=\cos\chi|00\rangle+\sin\chi|11\rangle$ and $\rho^{B}=\mathrm{tr}_{A}|\psi_{\chi}\rangle\langle\psi_\chi|$,
$0\leqslant\alpha\leqslant1,0<\chi\leqslant\frac{\pi}{4}$. It has been proved that $\rho(\alpha,\chi)$ is steerable from Alice to Bob if and only if the inequality $\cos^{2}2\chi\geqslant\frac{2\alpha-1}{(2-\alpha)\alpha^{3}}$ holds. That is, the state $\rho(\alpha,\chi)$ is a one-way steerable state. However, it is difficult to efficiently determine whether the arbitrary quantum state is steerable or not.

Subsequently, L. Vandenberghe and S. Boyd gave the numerical calculation quantum steering criterion in terms of SDP as follows\cite{SDP}.

Suppose that Alice performs $m$ measurements with $w$ outcomes each, i.e., $x=0,\ldots,m-1$ and $a=0,\ldots,w-1$, $\lambda^{'}$ is a function from $\{0,\ldots,m-1\}$ to $\{0,\ldots,w-1\}$. We can identify every $\lambda^{'}$ by a string of outcomes $\lambda^{'}=(a_{x=0},a_{x=1},\ldots,a_{x=m-1})$. Obviously, there are $d=w^m$ such strings. Define the deterministic probability distribution $D(a|x,\lambda^{'})$ as $D(a|x,\lambda^{'})=\delta_{a,\lambda^{'}(x)}$, hence there are $d$ such distributions. Then Eq. \eqref{2} can be rewritten as

\begin{align}\label{3}
\rho_{a|x}=\sum_{\lambda^{'}=1}^{d}D(a|x,\lambda^{'})\rho_{\lambda^{'}},
\end{align}
where $p(a|x,\lambda)=\sum_{\lambda^{'}=1}^{d}p(\lambda^{'}|\lambda)D(a|x,\lambda^{'})$, and $\rho_{\lambda^{'}}\equiv\int d\lambda p(\lambda)p(\lambda^{'}|\lambda)\sigma_{\lambda}$.

Now we write the SDP which determines that Alice can steer Bob \cite{SDP}. If the quantum state $\{\rho_{a|x}\}$ and the deterministic probability distribution $\{D(a|x,\lambda^{'})\}_{\lambda^{'}}$ are given, considering the minimum value of the objective funciton
\begin{equation}\label{4}
\min_{\{F_{a|x}\}} \mathrm{tr} \sum_{ax}F_{a|x}\rho_{a|x},
\end{equation}
where the Hermitian matrices $\{F_{a|x}\}$ are satisfy $\sum_{ax}F_{a|x}D(a|x,\lambda^{'})\geqslant 0 \quad\forall\lambda^{'}$ and $\mathrm{tr}\sum_{ax\lambda^{'}}F_{a|x}D(a|x,\lambda^{'})=1$. If the minimum value of the objective function \eqref{4} is negative, then quantum state $\rho$ is steerable from Alice to Bob. Whilst, if the minimum value of the objective function \eqref{4} is positive, then quantum state $\rho$ is unsteerable from Alice to Bob.


\subsection{BP artificial neural network model}
BP neural network usually refers to a multi-layer feedforward network trained by error BackPropagation Algorithm, and it is one of the most widely used neural network models. The term BackPropagation and its general use in neural networks was announced in \cite{BP1}, and its  modern overview is given in the textbook \cite{BP2}.

BP neural network has the ability of learning and storing a large amount of mapping relationships between inputs and outputs without knowing the mathematical equations of the relationships. A BP neural network model includes an input layer, several hidden layers and an output layer. In order to minimize the accumulated error of the network, its learning rule is to use the gradient descent method to continuously adjust the connection weights and thresholds (or biases) of the network through back propagation.

\begin{figure}
  \centering
  \includegraphics[width=3in]{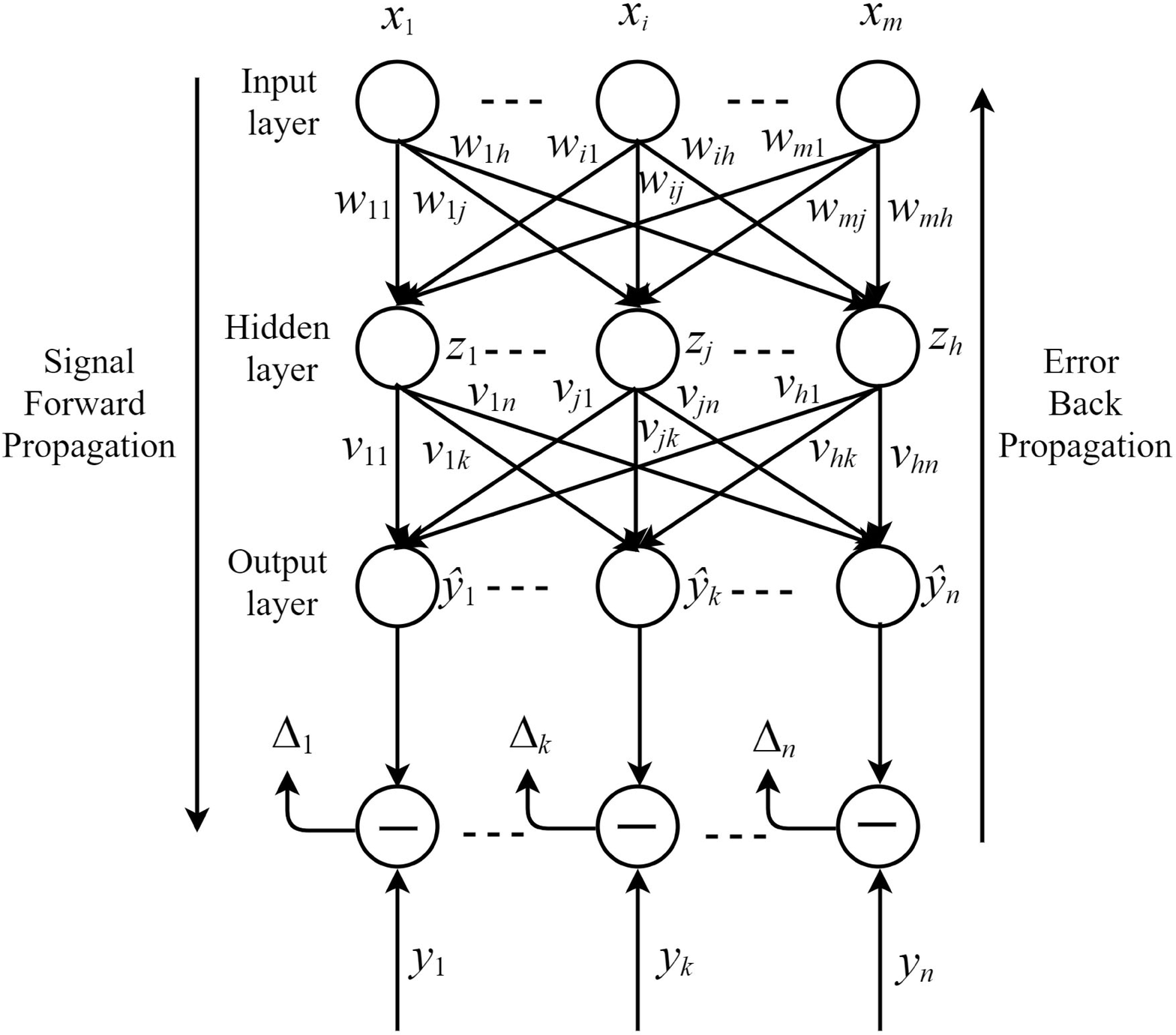}\\
  \caption{3-layer BP neural network model.}\label{fig1}
\end{figure}

For the convenience of discussion, we show a typical BP neutral network model with only 3 layers in Fig. \ref{fig1}. Given a training set $D=\{(\bm{x}_1,\bm{y}_1),(\bm{x}_2,\bm{y}_2),\ldots,(\bm{x}_d,\bm{y}_d)\}$, where $\bm{x}_t\in \mathbb{R}^m,\bm{y}_t\in \mathbb{R}^n$, the BP neutral network model has $m$ input-layer neurons, $n$ output-layer neurons and $h$ hidden-layer neurons, where the thresholds of the $j$th neuron in the hidden layer and the $k$th neuron in the output layer are denoted by $\gamma_j$ and $\theta_k$ respectively, the connection weight between the $i$th neuron in the input layer and the $j$th neuron in the hidden layer is denoted by $w_{ij}$, and the connection weight between the $j$th neuron in the hidden layer and the $k$th neuron in the output layer is labeled as $v_{jk}$. All the weights and thresholds are initialized by the Normal distribution.

By calculation, the input received by the $j$th neuron in the hidden layer is $\alpha_j=\sum_{i=1}^m w_{ij}x_i$, and the input received by the $k$th neuron in the output layer is $\beta_k=\sum_{j=1}^h v_{jk}z_j$. Suppose the activation functions from the input layer to the hidden layer and from the hidden layer to the output layer are $f$ and $g$ respectively and the output of the network is $\bm{\hat{y}}_t=(\hat{y}_1^t,\hat{y}_2^t,\ldots ,\hat{y}_n^t)$ for the training example $(\bm{x}_t,\bm{y}_t)$, the output of the $j$th neuron in the hidden layer is $z_j=f(\alpha_j-\gamma_j)$, and the output of the $k$th neuron in the output layer is $\hat{y}_k^t=g(\beta_k-\theta_k)$. In this paper, every activation function is chosen by the ReLU function

\begin{equation}
\mathrm{ReLU}(x)=\left\{\begin{aligned}
x,\quad \quad \mathrm{if}~x>0,\\
0,\quad \quad \mathrm{if}~x\leqslant0.
\end{aligned}\right.
\end{equation}

So the mean square error of the network on $(\bm{x}_t,\bm{y}_t)$ is

\begin{align}\label{bp1}
\begin{array}{l}
E_t=\frac{1}{2}\sum_{k=1}^n(\hat{y}_k^t-y_k^t)^2,
\end{array}
\end{align}
and the accumulated error (cost function) of the network on the training set $D$ is
\begin{align}\label{bp2}
\begin{array}{l}
E=\frac{1}{d}\sum_{t=1}^d E_t.
\end{array}
\end{align}
It should be noticed that the goal of the BP algorithm is to minimize the function (\ref{bp2}).

Now there are $(m+n+1)h+n$ parameters in the network of the Fig. \ref{fig1} to be determined: $mh$ connection weights between the input layer and the hidden layer, $hn$ connection weights between the hidden layer and the output layer, $h$ thresholds of the hidden layer neurons, and $n$ thresholds of the output layer neurons. BP algorithm is an iterative learning algorithm in which the parameters are updated and estimated in each round of iteration, and the updated formula of an arbitrary parameter $\omega$ is given by
\begin{align}\label{bp3}
\begin{array}{l}
\omega \leftarrow \omega +\Delta \omega.
\end{array}
\end{align}
BP algorithm adjust every parameter in the direction of the negative gradient of the target to minimize the cost function $E$ based on the gradient descent strategy. Given a learning rate $\eta$, any parameter $\omega$ can be adjusted $\eta$ units along the gradient descent direction every time, that is,
\begin{align}\label{bp4}
\begin{array}{l}
\Delta\omega=-\eta\frac{\partial E}{\partial\omega}.
\end{array}
\end{align}
After continuous iterations, the error $E$ can be minimized.

In summary, we show the workflow of the BP algorithm in Table \ref{table}. For each training example, the BP algorithm performs the following operations. First of all, the algorithm provides the input example to the input-layer neurons, and then propagates forward the signal layer by layer until the output layer generates a result; Then the network calculates the error of the output layer (lines 4-5), propagate the error back to the hidden-layer neurons (line 6), and adjust the connection weights and thresholds according to the error of the hidden-layer neurons (line 7). Finally, this iterative process loops until the cost function reaches the minimum.

\begin{table}[htbp]
  \centering
    \begin{tabular}{ll}
\hline
    \textbf{Input:} &  Training set $D=\{(\bm{x}_t,\bm{y}_t)\}_{t=1}^d$; \\
          & Learning rate $\eta$. \\
    \textbf{Process:}&  \\
  \multicolumn{2}{l}{~~1: Randomly initialize all connection weights and threshol-}\\
  \multicolumn{2}{l}{ds in the network within the range of (0,1)}\\
  \multicolumn{2}{l}{~~2: \textbf{repeat}}\\
  \multicolumn{2}{l}{~~3: \quad \textbf{for all} $(\bm{x}_t,\bm{y}_t)\in D$ \textbf{do}} \\
  \multicolumn{2}{l}{~~4: \quad\quad Calculate the current output $\hat{\bm{y}}_t$;}\\
  \multicolumn{2}{l}{~~5: \quad\quad Calculate the gradient term of the neurons in the o-}\\
  \multicolumn{2}{l}{utput layer;} \\
  \multicolumn{2}{l}{~~6: \quad\quad Calculate the gradient term of the neurons in the h-}\\
  \multicolumn{2}{l}{idden layer;} \\
  \multicolumn{2}{l}{~~7: \quad\quad Update connection rights $w_{ij}, v_{jk}$ and biases $\gamma_j,\theta_k$}\\
  \multicolumn{2}{l}{~~8: \quad\textbf{end for}} \\
  \multicolumn{2}{l}{~~9: \textbf{until} the cost function reaches the minimum} \\
    \textbf{Output:} &  Multi-layer feedforward neural network determined\\
  &  by connection weights and thresholds. \\
\hline
    \end{tabular}%
  \caption{Error BackPropagation Algorithm}
  \label{table}%
\end{table}
\section{Detecting the steerability by BP neural networks}\label{sec}
SVM maps the samples from the original space to a higher-dimensional feature space through the kernel function, so that the samples are linearly separable in this feature space and the global optimal solution can be found. But the data generated by SDP inevitably exists errors, so the models trained by SVM have errors. Because of the advantage of easily adjusting the parameters and a better performance when processing large amounts of data and continuous data features of the neural network, we use the BP neural network to detect the quantum steerability.

\subsection{Datasets}
In order to train the quantum steering classifiers, we need to collect the data of quantum states and select the features for the data. Inspired by Ref. \cite{30}, we generate two random $4 \times 4$ complex matrices $A$ and $B$. Then we use the two matrices to generate a Hermitian matrix $H\equiv(A+\mathrm{i}B)(A+\mathrm{i}B)^{\dagger}$, where $\dagger$ means conjugate transpose. Finally, the density matrix $\rho\equiv H/\mathrm{tr}(H)$ can be obtained. Thus, utilizing SDP to find out whether the each sample quantum state is steerable or not whilst label the quantum state to be ``$-1$" and ``$+1$" , respectively. In this paper, we use the datasets collected in \cite{30}, for which we select the features in the following four cases.

F1: Every feature vector is composed of 15 features in F1, that is, $\rho_{ii},i\in\{1,2,3\}$, the real and imaginary part of $\rho_{ij},i>j$.

F2: Every feature vector is composed of 9 features in F2, that is, $\mathrm{tr}[(\sigma_{k} \otimes \sigma_{l})\rho],\{k, l\}\in \{1,2,3\}$.

An arbitrary two-qubit density operator $\rho$ in Bloch representation can be written as

\begin{align}
\rho=&\frac{1}{4}(I+\sum_{i=1}^{3} x_{i} \sigma_{i} \otimes I^{B}+\sum_{j=1}^{3} y_ {j} I^{A} \otimes \sigma_{j}+\notag
\\&\sum_{k,l=1}^{3} t_{kl} \sigma_{k} \otimes \sigma_{l}),
\end{align}
where $t_{kl}=\mathrm{tr}[(\sigma_{k}\otimes\sigma_{l})\rho]$ can construe the correlation matrix that stands for a certain quantum correlation. The partial information is extracted by computing $\mathrm{tr}[(\sigma_{k}\otimes\sigma_{l})\rho]$ as features. If the machine learning model with high performance can be trained by selecting these 9 features, we can judge whether an arbitrary unknown two-qubit quantum state is steerable by only measuring in three fixed directions $x,y,z$.

F3: Every feature vector is composed of 9 features in F3, that is, $\rho \to \rho'\equiv(I^{A}\otimes\sqrt{\rho^{B}})\rho (I^{A}\otimes\sqrt{\rho^{B}}),\mathrm{tr}[(\sigma_{k}\otimes\sigma_{l})\rho'],\{k,l\}\in\{1,2,3\}$.

To further explore a high-performance machine learning model with partial information, we convert the state $\rho$ into a canonical form $\rho'$ by local unitaries, which preserves the steerability of $\rho$. As proved in Ref. \cite{31}, the map is given by
\begin{align}
\rho \to \rho'\equiv(I^{A}\otimes\sqrt{\rho^{B}})\rho(I^{A}\otimes\sqrt{\rho^{B}}),
\end{align}
where $\rho^{B}=\mathrm{tr}_{A}\rho$.

Similarly, we extract the coefficients of the correlation terms of $\rho'$, $t'_{kl}$ to combine the feature vector. Same to the case of F2, we only need to measure an arbitrary two-qubit state in three fixed directions: $x,y,z$ to predict the steerability of it.

F4: Every feature vector is composed of 6 features in F4, that is, $\mathrm{F}_{3}$ except for the terms of $\{\sigma_{2}\otimes\sigma_{1},\sigma_{3}\otimes\sigma_{1},\sigma_{3}\otimes\sigma_{2}\}$.

To explore a high performance-machine learning model of steering with less information, according to the symmetry, we drop the coefficients of the correlation terms, $\{\sigma_{2}\otimes\sigma_{1},\sigma_{3}\otimes\sigma_{1},\sigma_{3}\otimes\sigma_{2}\}$.

According to the numbers of measurements $m=2,3,\ldots,8$, the whole dataset can be divide into 28 datasets after selecting the feature vectors. For each $m$, there exists at least 5000 examples with the label ``$+1$" and 5000 examples with the label ``$-1$" in the corresponding dataset. From the data sets, we can select 1000 positive examples and 1000 negative examples as the test set randomly, and the residual as the training set. We employ the BP neural network to train a model for every dataset.

\subsection{Training and testing}
The neural networks used in this article are all have 2 hidden layers, and every layer has 200-1000 neurons. We choose ReLU as the activation function and Adam for the gradient descent method. The batch sizes of the models are all 200 and learning rate equals 0.001.

\begin{figure}
  \centering
  \includegraphics[width=3.3in]{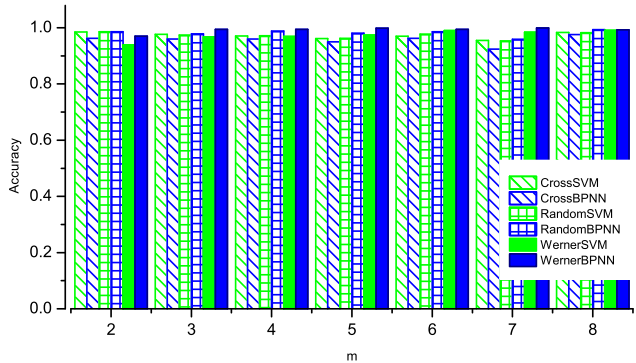}\\
  \caption{Classification accuracy of machine learning with F1 features. The first and second columns (diagonal-filled green and diagonal-filled blue) depict the accuracy of cross validation of the classifiers trained by SVM and BP neutral network, respectively. The third and forth columns (grid-filled green and grid-filled blue) depict the classification accuracy on random states of the classifiers trained by SVM and BP neutral network, respectively. The fifth and sixth columns (solid green and solid blue) depict the classification accuracy on the Werner state of the classifiers trained by SVM and BP neutral network, respectively.}\label{fig3}
\end{figure}

\begin{figure}
  \centering
  \includegraphics[width=3.3in]{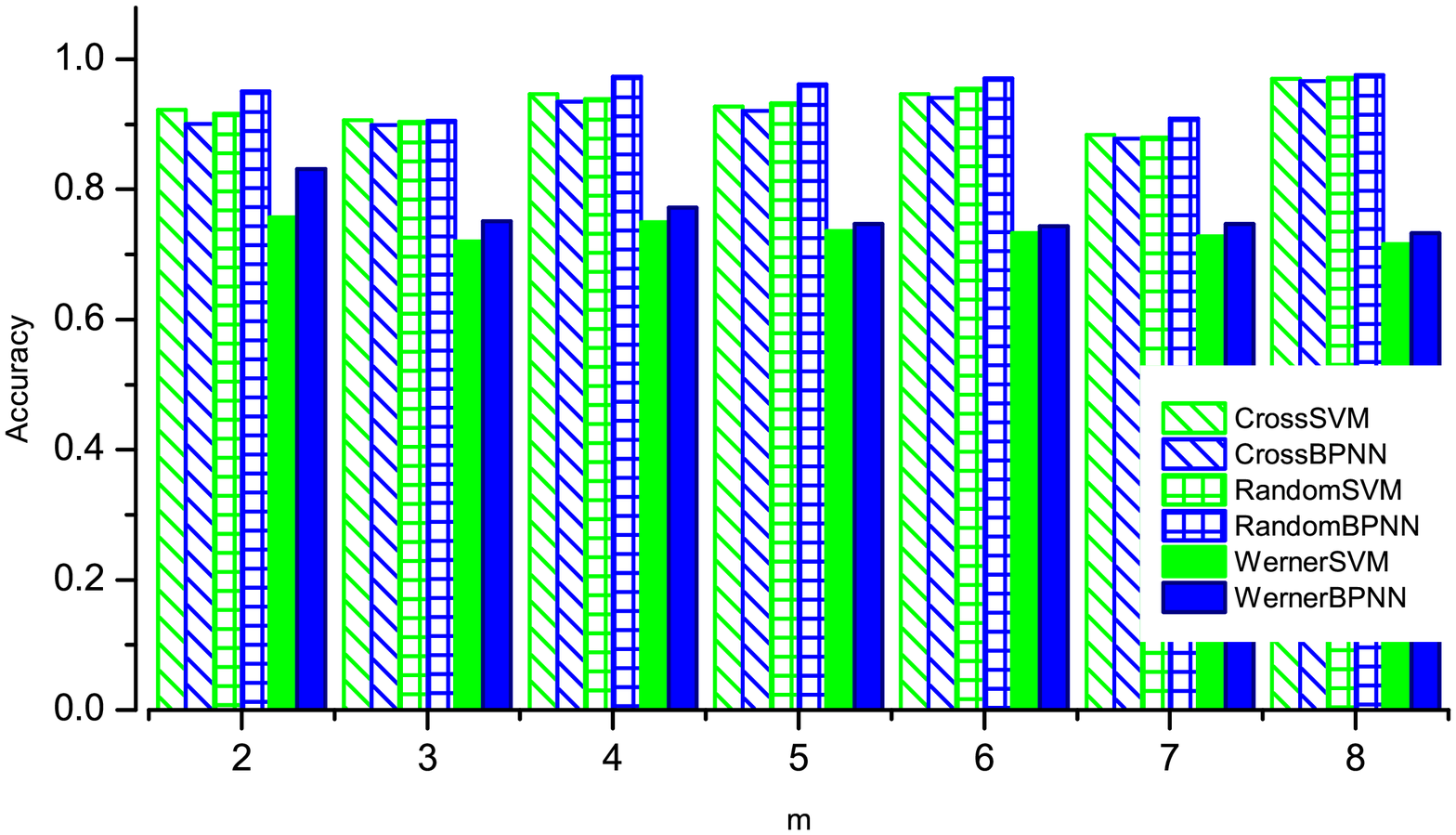}\\
  \caption{Classification accuracy of machine learning with F2 features. The first and second columns (diagonal-filled green and diagonal-filled blue) depict the accuracy of cross validation of the classifiers trained by SVM and BP neutral network, respectively. The third and forth columns (grid-filled green and grid-filled blue) depict the classification accuracy on random states of the classifiers trained by SVM and BP neutral network, respectively. The fifth and sixth columns (solid green and solid blue) depict the classification accuracy on the Werner state of the classifiers trained by SVM and BP neutral network, respectively.}\label{fig4}
\end{figure}

\begin{figure}
  \centering
  \includegraphics[width=3.3in]{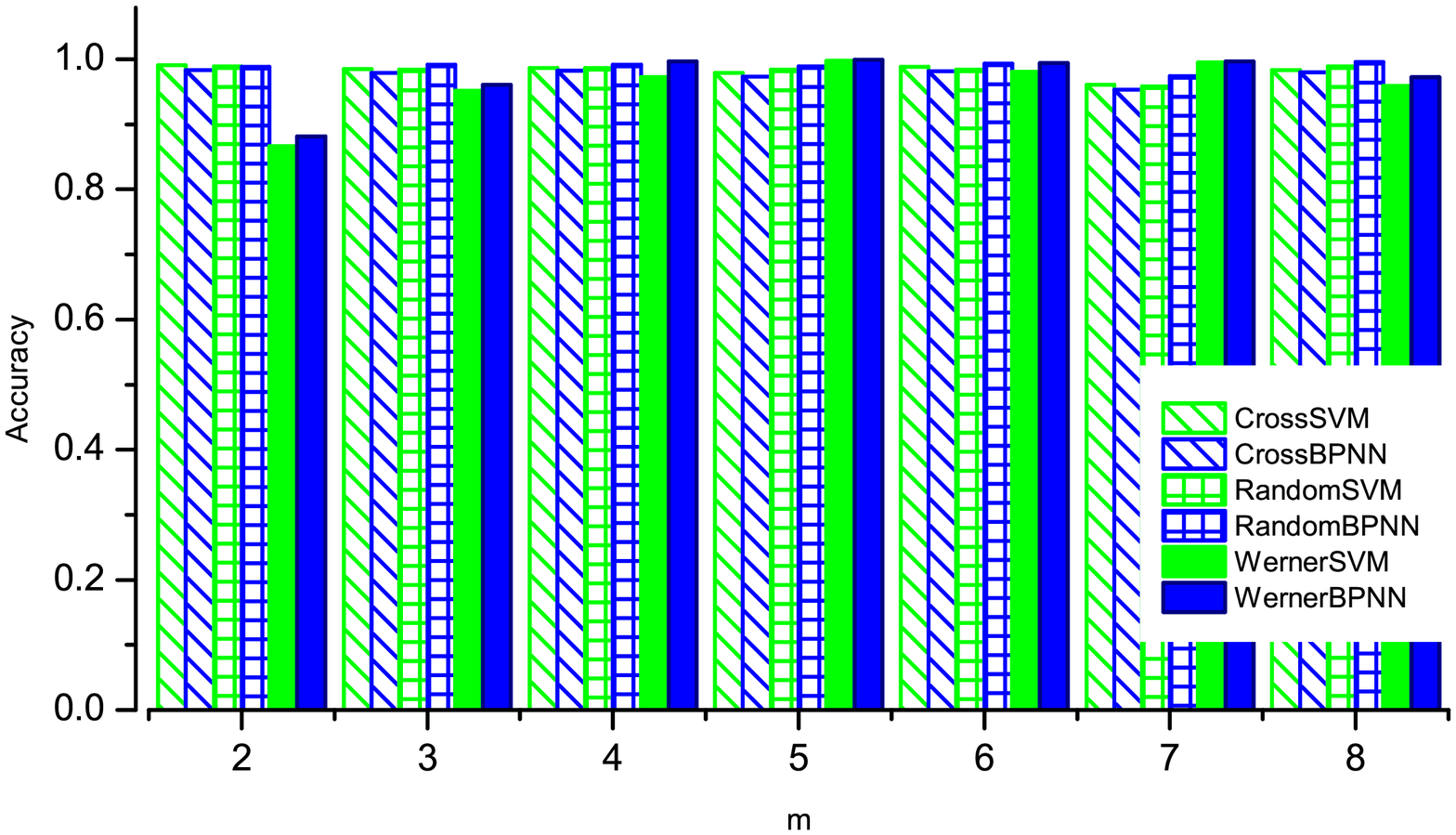}\\
  \caption{Classification accuracy of machine learning with F3 features. The first and second columns (diagonal-filled green and diagonal-filled blue) depict the accuracy of cross validation of the classifiers trained by SVM and BP neutral network, respectively. The third and forth columns (grid-filled green and grid-filled blue) depict the classification accuracy on random states of the classifiers trained by SVM and BP neutral network, respectively. The fifth and sixth columns (solid green and solid blue) depict the classification accuracy on the Werner state of the classifiers trained by SVM and BP neutral network, respectively.}\label{fig5}
\end{figure}

\begin{figure}
  \centering
  \includegraphics[width=3.3in]{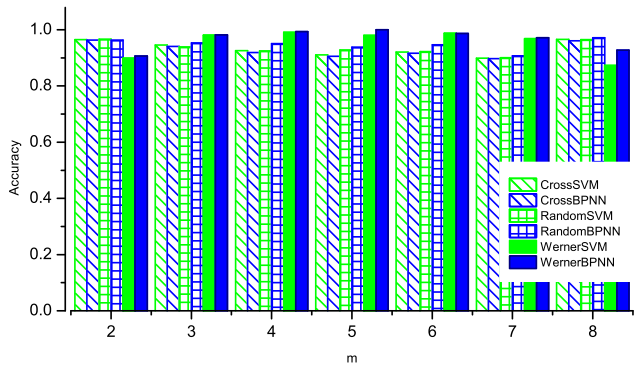}\\
  \caption{Classification accuracy of machine learning with F4 features. The first and second columns (diagonal-filled green and diagonal-filled blue) depict the accuracy of cross validation of the classifiers trained by SVM and BP neutral network, respectively. The third and forth columns (grid-filled green and grid-filled blue) depict the classification accuracy on random states of the classifiers trained by SVM and BP neutral network, respectively. The fifth and sixth columns (solid green and solid blue) depict the classification accuracy on the Werner state of the classifiers trained by SVM and BP neutral network, respectively.}\label{fig6}
\end{figure}

After a neural network model is trained, we test the performance by creating a set of feature vectors of new quantum states, which is different from the dataset used for training. For a test dataset with known labels, the classification accuracy of the learned model on the test set is the percentage of the number of examples correctly predicted to the size of the test set. The Fig. \ref{fig3},\ref{fig4},\ref{fig5},\ref{fig6} represent classification accuracies of machine learning models with F1,F2,F3,F4 features, respectively. In the four figures, we select every test set as a subset of the training set or a new dataset different from the training set to demonstrate the generalization ability and compare the three kinds of accuracies of the classifiers trained by BP neural network with that trained by SVM \cite{30}.


Firstly, in order to verify whether the model is well trained, we select a subset of the training set as the test set. Inspired by Ref. \cite{30}, in our experiments, we take the method of fourfold cross validation. Namely, we divide the entire training set into four equal parts randomly and take each part in turn as the test set, and the rest as the training set. Based on the trained four classifiers, we can obtain the accuracy of cross validation which is the average accuracy of the four classifiers. This accuracy information is used to generate the final classifier. In Fig. \ref{fig3},\ref{fig4},\ref{fig5},\ref{fig6}, the final cross validation accuracies of the classifiers trained by SVM are illustrated by the first (diagonal-filled green) columns, and that trained by BP neural network are illustrated by the second (diagonal-filled blue) columns.

Secondly, the classification accuracies of the models trained by SVM and BP neural network on the random test set, formed by the reserved 2000 examples, are illustrated by the third (grid-filled green) and the forth (grid-filled blue) columns respectively. All accuracies are higher than 0.9, which clearly shows that the models are well trained.

Finally, the last classification accuracies of the models trained by SVM and BP neural network on the test set generated by the Werner state ($\chi=\frac{\pi }{4}$) which are illustrated as solid green and solid blue columns respectively in Fig. \ref{fig3},\ref{fig4},\ref{fig5},\ref{fig6}.

As we know, the generalized Werner states is steerable from Alice to Bob if the following condition holds,
\begin{align}\label{11}
\cos^2 2\chi\geqslant\frac{2\alpha-1}{(2-\alpha)\alpha^3}.
\end{align}
Obviously, the bound of the parameter $\alpha$ that Alice can steer Bob's state is determined by Eq. (\ref{11}). We now construct generalized Werner state based on the uniform distribution of $\alpha$ and $\chi$. For each $\chi =\{\frac{\pi }{4},\frac{\pi }{6},\frac{\pi }{8},\frac{\pi }{12}\} $, we can generate 10000 states and use them to create a dataset for every feature (F1,F2,F3,F4), where each example has a feature F$i$($i$=1,2,3,4), and a label ($+1$ for steerable and $-1$ for unsteerable) determined by Eq. (\ref{11}). Notice that these states have completely correct labels.

Interestingly, for all features, the classification accuracies of the Werner state of the models trained by BP neural network are all higher than that training by SVM. Some accuracies of fourfold cross validation and on the random datasets of the models trained by BP neural network have increased, and even if some other accuracies have decreased but not very much. This shows that our models trained by BP neural network have good generalization abilities. This is because some unsteerable states computed by SDP may be misjudged. The reason is that the number of measurement values reaches 100 and fails to return a negative objective value of Eq. (\ref{4}), the SDP is stopped and marked that the quantum state is not steerable; in other word, the number of measurement values is increased, negative values may be returned and the quantum state can be labeled as a steerable state.

\subsection{Optimizing the steerability bounds}
In this section, we use the quantum steering classifiers that trained by BP neutral network to predict the steerability bounds of the generalized Werner state, and compare them with the bounds that computed by SDP and trained by SVM \cite{30}.

\begin{figure}
  \centering
  \includegraphics[width=3.5in]{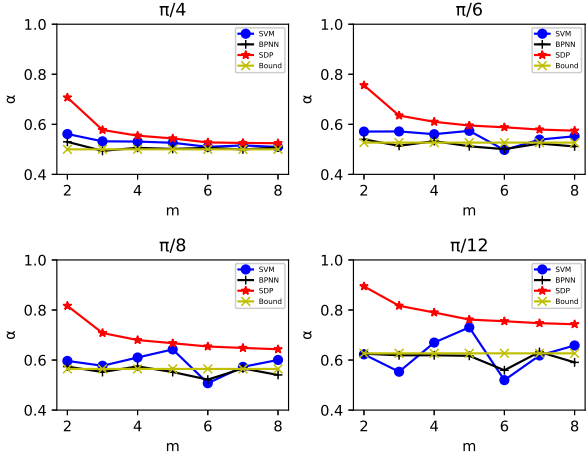}\\
  \caption{The predictions of steerability for generalized Werner states by learned classifiers and SDP. The blue-circle line is the result predicted by learning classifiers trained by SVM with F1 features, the black-plus line is the result predicted by learning classifiers trained by BP neural network with F1 features, the red-star line is the result predicted by SDP, and the yellow-cross line is the steerability bound from Alice to Bob.}\label{fig7}
\end{figure}

\begin{figure}
  \centering
  \includegraphics[width=3.5in]{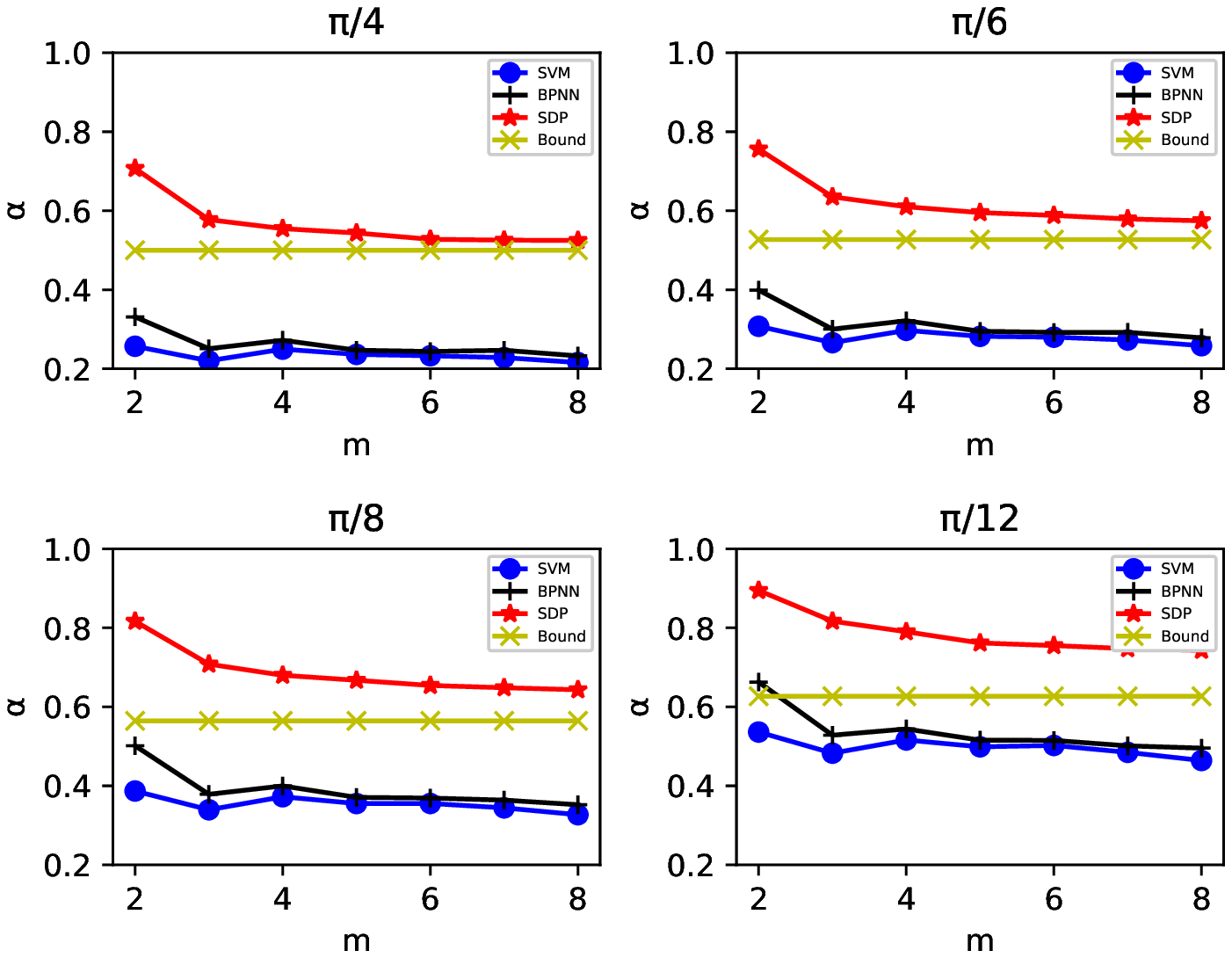}\\
  \caption{The predictions of steerability for generalized Werner states by learned classifiers and SDP. The blue-circle line is the result predicted by learning classifiers trained by SVM with F2 features, the black-plus line is the result predicted by learning classifiers trained by BP neural network with F2 features, the red-star line is the result predicted by SDP, and the yellow-cross line is the steerability bound from Alice to Bob.}\label{fig8}
\end{figure}

\begin{figure}
  \centering
  \includegraphics[width=3.5in]{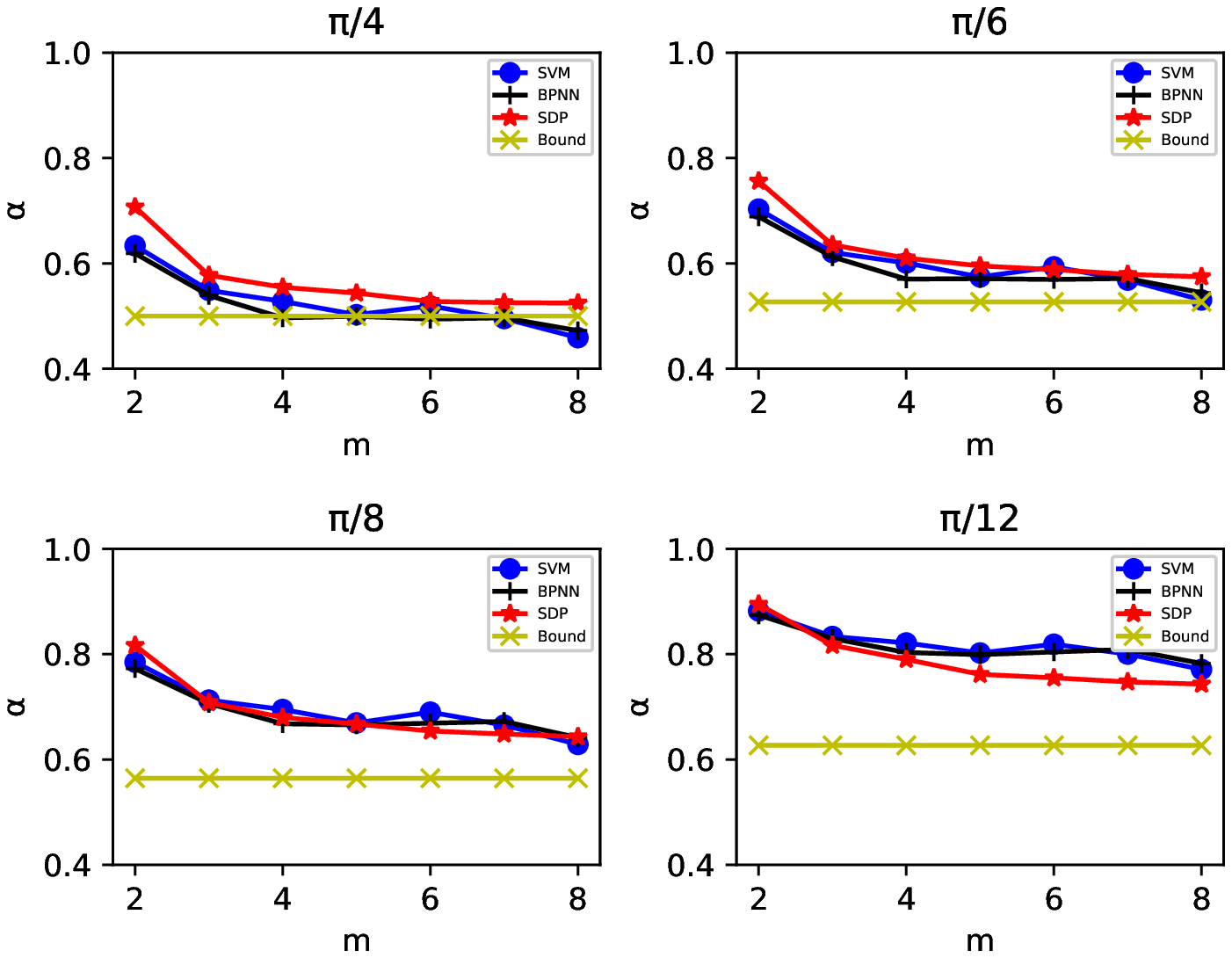}\\
  \caption{The predictions of steerability for generalized Werner states by learned classifiers and SDP. The blue-circle line is the result predicted by learning classifiers trained by SVM with F3 features, the black-plus line is the result predicted by learning classifiers trained by BP neural network with F3 features, the red-star line is the result predicted by SDP, and the yellow-cross line is the steerability bound from Alice to Bob.}\label{fig9}
\end{figure}

\begin{figure}
  \centering
  \includegraphics[width=3.5in]{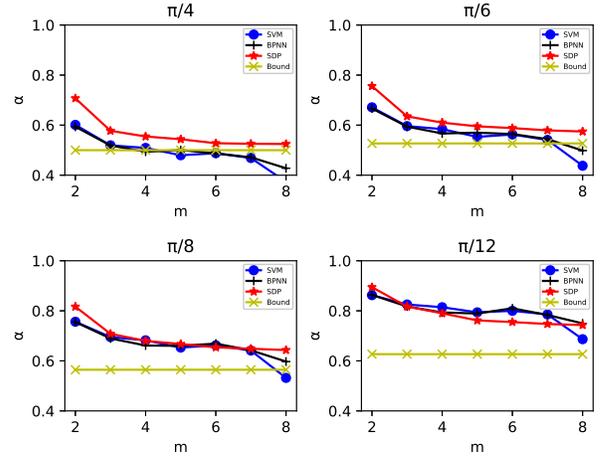}\\
  \caption{The predictions of steerability for generalized Werner states by learned classifiers and SDP. The blue-circle line is the result predicted by learning classifiers trained by SVM with F4 features, the black-plus line is the result predicted by learning classifiers trained by BP neural network with F4 features, the red-star line is the result predicted by SDP, and the yellow-cross line is the steerability bound from Alice to Bob.}\label{fig10}
\end{figure}

As shown in Fig. \ref{fig7},\ref{fig8},\ref{fig9},\ref{fig10}, these four figures correspond to the classifiers trained with features F1, F2, F3, and F4, respectively. As illustrated in Fig. \ref{fig7}, the four subfigs in this figure correspond to $\chi=\{\frac{\pi}{4},\frac{\pi}{6},\frac{\pi}{8},\frac{\pi}{12}\}$ respectively. In each subfig, the blue-circle lines stand for the results predicted by the classifiers trained by SVM for $m=2,\ldots,8$, respectively. The black-plus lines are the results predicted by the classifiers trained by BP neutral network for $m=2,\ldots,8$, respectively. The red-star lines are the results predicted by SDP with $m=2,\ldots,8$. The yellow-cross lines are the steerability bounds from Alice to Bob which are determined by Eq. (\ref{11}). Notice that the steering bounds which are predicted by SVM (blue-circle lines) and BP neural network (black-plus lines) are all lower than the bounds calculated by SDP (red-star lines). Especially, when $\chi=\frac{\pi}{4}$, the generalized Werner states reduces to the Werner state, the bounds predicted by classifiers are always higher than the theoretical bounds (yellow-cross lines) and the bounds predicted by BP neural network are lower than the bounds predicted by SVM and SDP, namely, the bounds are more closer to the theoretical bounds. In addition, as illustrated in Fig. \ref{fig8},\ref{fig9},\ref{fig10}, they show that the bounds predicted by BP neural network are obviously very close to the theoretical bounds.


During the process of the predicted the bounds of the generalized Werner states, comparing with the bounds predicted by SVM, the bounds predicted by our models are more closer to the theoretical bounds, although a few bounds predictions are lower than the theoretical bounds. This shows that the learning classifiers may be more better than SDP, but they still have the possibilities of predicting steerability bounds lower than the theoretical bounds, which almost never happens for SDP. This is because the error of SDP mainly occurs when it finds out the data sets. The error date sets lead to the machine learning predict positive to be negative, and predict negative to be positive. With the decrease of $\chi$, the prediction errors of the learning classifiers and SDP are all increase, because of the predictions of the marginal states become more and more difficult. As shown in Fig. \ref{fig7},\ref{fig8},\ref{fig9},\ref{fig10}, the predicted bound made by classifiers that trained by BP neutral network are remain to be better than that made by the SVM method.

The above results clearly show that the BP neural network is effective in steerability detection. Compared with the SVM method in Ref. \cite{30}, our method makes not only none significantly reduction of the accuracies of the classifiers, but also the predictions of the bounds of the generalized Werner states closer to the theoretical bounds. Moreover, our method makes the time-consuming roughly the same as the SVM method. Now we take $m=8$ as an example, the learning classifiers trained by these two kinds of machine learning methods take about $10^{-2}$ seconds to predict an unknown state, while for the SDP program it takes about $10^2$ seconds. This also illustrates the time advantage of machine learning classifiers.

\subsection{Comparing the classifiers trained with four features}
From Fig. \ref{fig4} to Fig. \ref{fig6}, we explore classifiers with partial information of a quantum state with F2, F3 and F4 features. Due to the accuracies in Fig. \ref{fig4} are lower than others, it is clear that F3 and F4 are all better than F2. Next, let's compare the performance of classifiers trained with features F$i, i=1,2,3,4$ on different datasets.

\begin{figure}
  \centering
  \includegraphics[width=3.5in]{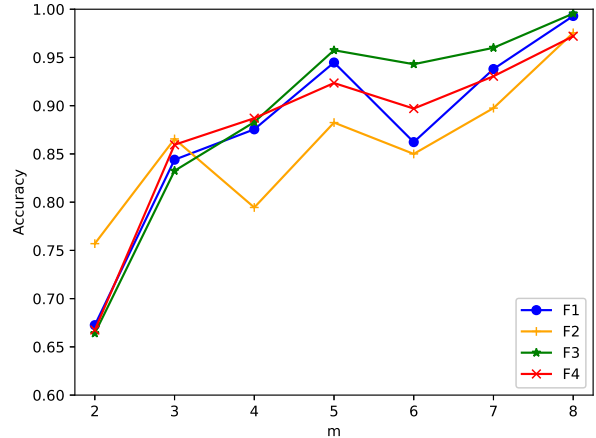}\\
  \caption{Accuracy of classification on random test data of $m=8$ with different classifiers.}\label{fig11}
\end{figure}

As we know, with the measurement $m$ increase, SDP could find out the quantum state steerable or unsteerable more accurate. Thus, it is reasonably that the steerability classifiers are well trained with the measurement $m$ increases. Next, we prove the validity of the classifiers trained by the data sets for $m=8$, and use these classifiers to test the data sets for different $m$. In Fig. \ref{fig11}, the blue-circle line, the orange-plus line, the green-star line, the red-cross line stand for different features F1, F2, F3 and F4, respectively. It is obvious that, the trend of test accuracy are grow rapidly and it is generally the same as the theoretical prediction. However, the curves occur inflection points at $m=6$ and there exists the second inflection point for steerability classifier of feature F2 at $m=4$. This phenomenon comes from the drawback in the training process. In the whole, though the range of accuracies is relatively large, the their accuracies tendency is correct. And the steerability classifier of feature F3 presents the higher performance.

\begin{figure}
  \centering
  \includegraphics[width=3.5in]{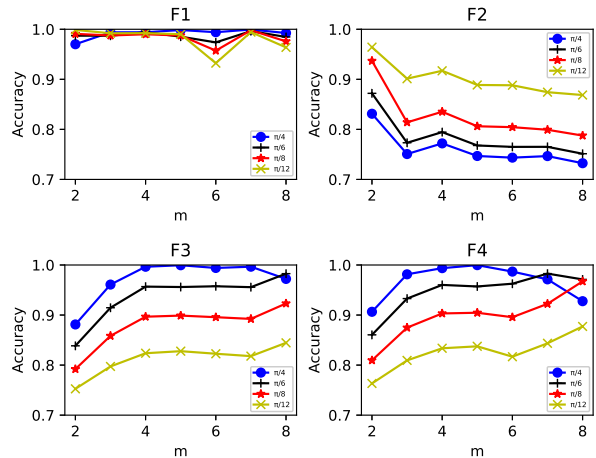}\\
  \caption{Classification accuracy for generalized Werner states by each machine learing classifier with the different features: F1, F2, F3, and F4, respectively.}\label{fig12}
\end{figure}

In Fig. \ref{fig12}, it shows that the classification accuracy for the generalized Werner states ($\chi=\frac{\pi}{4},\frac{\pi}{6},\frac{\pi}{8},\frac{\pi}{12}$) by each machine learing classifier with the different features F1, F2, F3, and F4, respectively. The classification accuracy of the generalized Werner states tested by the steerability classifier of feature F1 presents in the first subfig, it maintains at a high level. With the parameter $\chi$ decreases, the accuracy will be decrease.
In the second subfig of the steerability classifier of feature F2, the trend of curves grow down, it demonstrates that the ability of predicted the bounds of the generalized Werner states is poor. The third and forth subfigs show that the trend of the accuracy have the similar trends. They illustrate that the more measurements, the higher accuracy of the classification and the accuracy decreases with decreasing parameters $\chi$. That is, the steerability classifiers of features F3 and F4 present the relative higher performance of predicted the bounds of the generalized Werner states.

\section{Conclusion}\label{sec4}

In this work, we have applied the BP neural network to construct several classifiers to identify the steerability of the two-qubit quantum states and optimize the steerability bounds of the generalized Werner states. Our main purpose is to construct machine learning models with stronger generalization abilities. Firstly, we find out that no matter what the features (F1,F2,F3,F4) of the quantum states are, the BP neural network can construct several models to realize high-performance quantum steering classifiers compared with the SVM approach. Secondly, we use the BP neural network to construct several new classifiers to predict the steerability bounds of the generalized Werner states, which shows that the predictions of the steerability bounds are closer to the theoretical bounds, i.e., it is very effective for testing or predicting the steerability of a large amount of the generalized Werner states. Finally, we particularly construct high-performance classifiers with partial information which we only need to measure in three fixed measurement directions to detect the steerability of arbitrary states validly. In conclusion, it shows that BP neural network can be very effective for identifying the steerability of a large amount of arbitrary states in the quantum information processing.

\emph{Acknowledgement.}---This work was supported by the National Natural Science Foundation of China (Grant No. 11771011), the Natural Science Foundation of Shan-xi Province, China (Grant No. 201801D221032, 201801D 121016) and Scientific and Technological Innovation Programs of Higher Education Institutions in Shanxi (Grant No. 2019L0178).

\end{document}